\documentstyle[11pt,newpasp,twoside,epsf]{article}
\begin{document}            

\title{The RMS Survey: A Systematic Search for Massive Young Stars in the Galaxy}
\author{M. G. Hoare, S. L. Lumsden, R. D. Oudmaijer, A. L. Busfield and T. L. King}
\affil{Department of Physics and Astronomy, University of Leeds, Leeds, LS2 9JT, UK}
\author{T. L. J. Moore}
\affil{Astrophysics Research Institute, Liverpool John Moores University, Twelve Quays House,
Egerton Wharf, Birkenhead, CH41 1LD, UK}

\begin{abstract}

We have selected red MSX sources (RMS) that have the colours of
massive young stellar objects (MYSOs). Our aim is to generate a large,
systematically selected sample to address questions such as their
luminosity function, lifetimes, clustering and triggering. Other
objects such as UCHIIs, PN, PPN and AGB stars have similar IR colours
and a large programme of ground-based follow-up observations is
underway to identify and eliminate these from the sample of the red
MSX sources. These include radio continuum observations, kinematic
distances, ground-based mid-IR imaging, near-IR imaging and
spectroscopy to distinguish. We report the progress of these campaigns
on the 3000 candidates, with initial indications showing that a
substantial fraction are indeed massive YSOs.

\end{abstract}

\section{Introduction}

Studies of massive star formation are currently hampered by the lack of
large systematically selected samples of objects in the earliest phases.
Any well-selected sample of massive YSOs must start from the IR where
most of their bolometric luminosity emerges.  IRAS data has previously been used
(e.g. Campbell et al. 1989; Chan et al. 1996; Sridharan et al. 2002), 
but its large beam size means that it is often confused in
the galactic plane. The MSX survey of the plane in the mid-IR (Price
et al. 2001) provides much better spatial resolution and virtually
eliminates any bias against the dense and clustered environments where
MYSOs are.  

We have developed colour-cuts ($F_{21}>2F_{8}$ and
$F_{8}<F_{14}<F_{21}$) to select MYSO candidates from the MSX PSC with
additional colour-cuts ($F_{8}>5F_{K}$ and $F_{K}>2F_{J}$) using the
2MASS near-IR survey (Lumsden et al. 2002).  These colour-cuts and the
elimination of known sources leaves a sample of about 1700 (leaving
out the crowded galactic centre region). This sample still contains
many other objects such as compact H II regions, PN, PPN, AGB stars,
etc., which have the same IR colours. Our follow-up programme of
observations is designed to identify and eliminate these.

\section{The RMS Survey Follow-Up Programme}

\begin{figure}
\plottwo{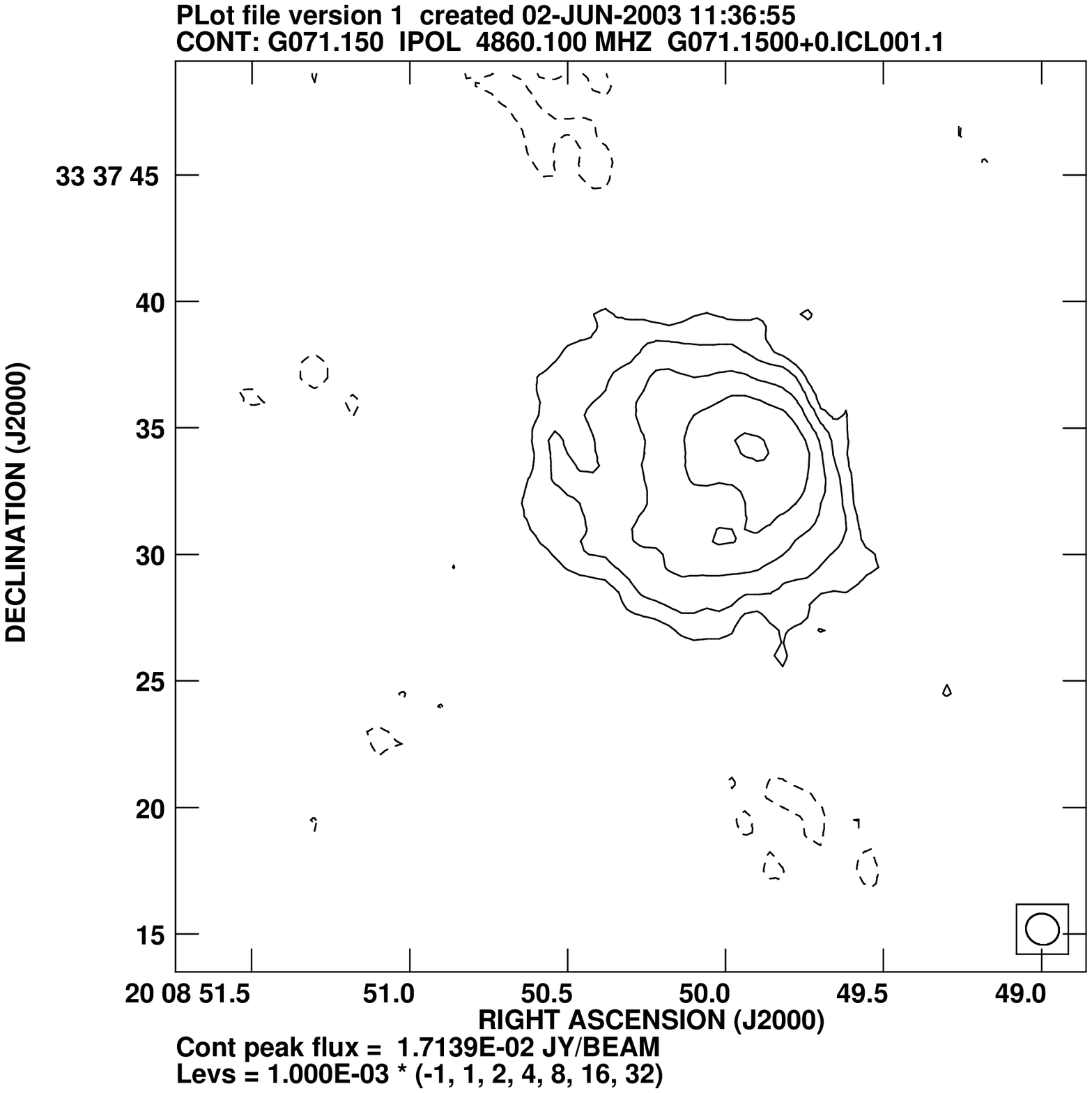}{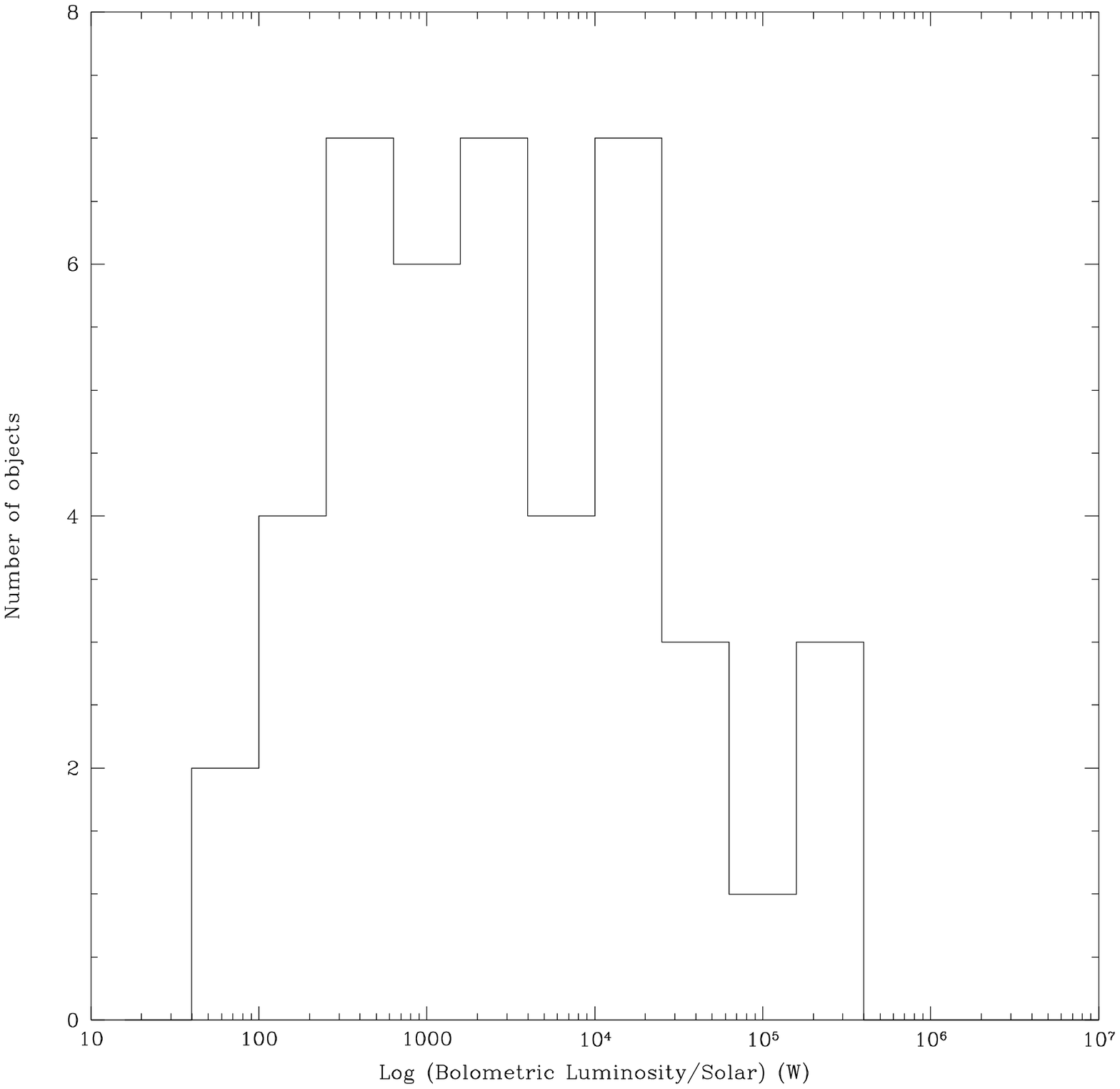}
\caption{a) Example of a compact, cometary H II region found in the RMS sample. 
b) Luminosity distribution of a sample of outer
galaxy RMS sources with good IRAS fluxes.}
\end{figure}

The main overlap is with compact H II regions, which are of course
also young massive stars in the phase immediately following the MYSO
phase.  These (and PN) are most easily identified by their strong
radio continuum emission compared to the weak emission from the
stellar winds of MYSOs (Hoare 2002). We are therefore observing every
candidate at 1 \arcsec\ resolution at 5 GHz. 500 targets have been
observed to date with about 20\% of objects detected, e.g. Figure 1a.
The remaining radio-quiet objects are kept as MYSO candidates.

To determine luminosities we are obtaining kinematic distances from
observations of $^{13}$CO and have so far observed 500 sources. Figure
1b is the luminosity distribution for a sub-sample of outer galaxy
sources, which clearly shows a substantial fraction of the sources
have luminosities consistent with early-type stars.

\begin{figure}
\vspace*{-2cm}
\plotfiddle{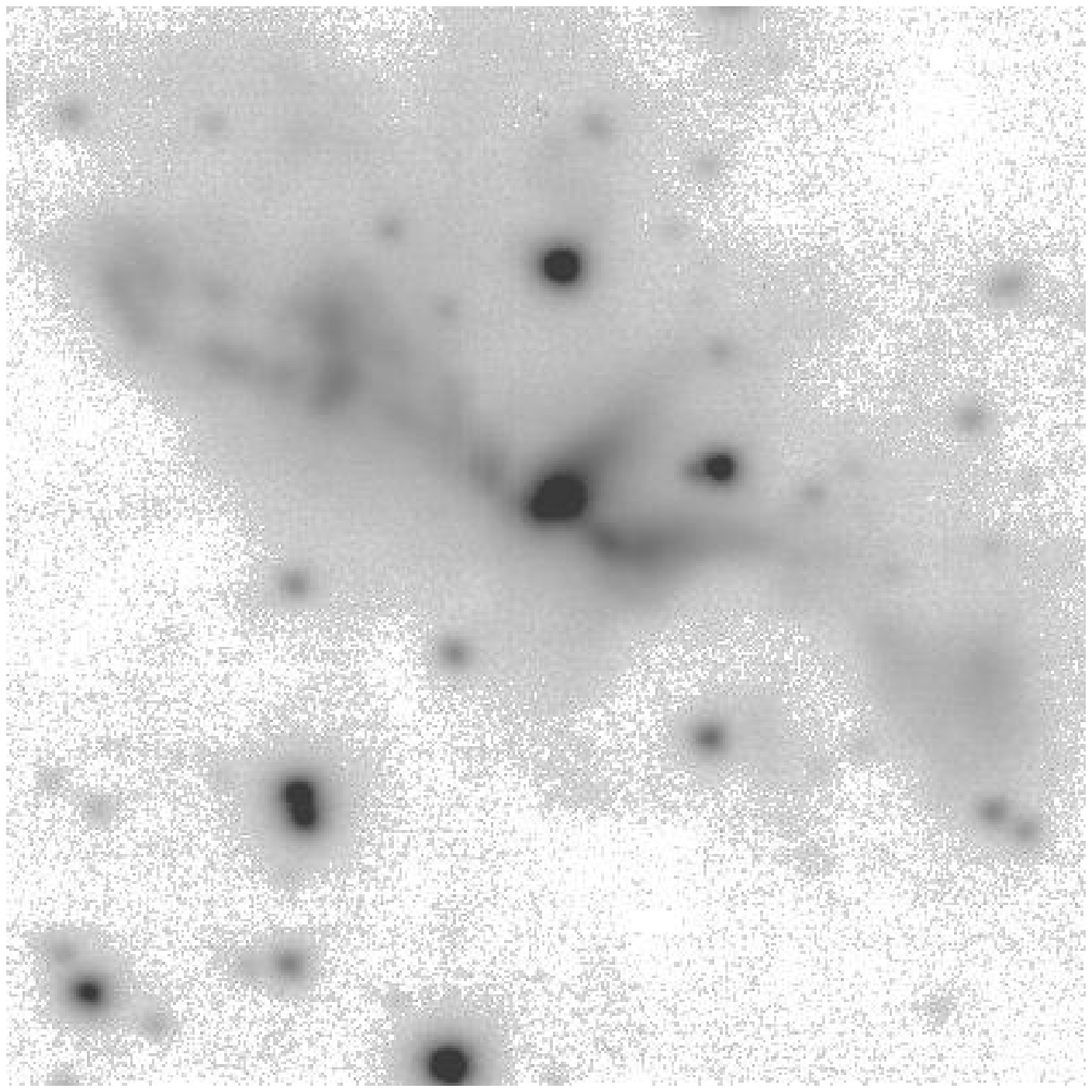}{3cm}{0}{30}{30}{-200}{-150}
\plotfiddle{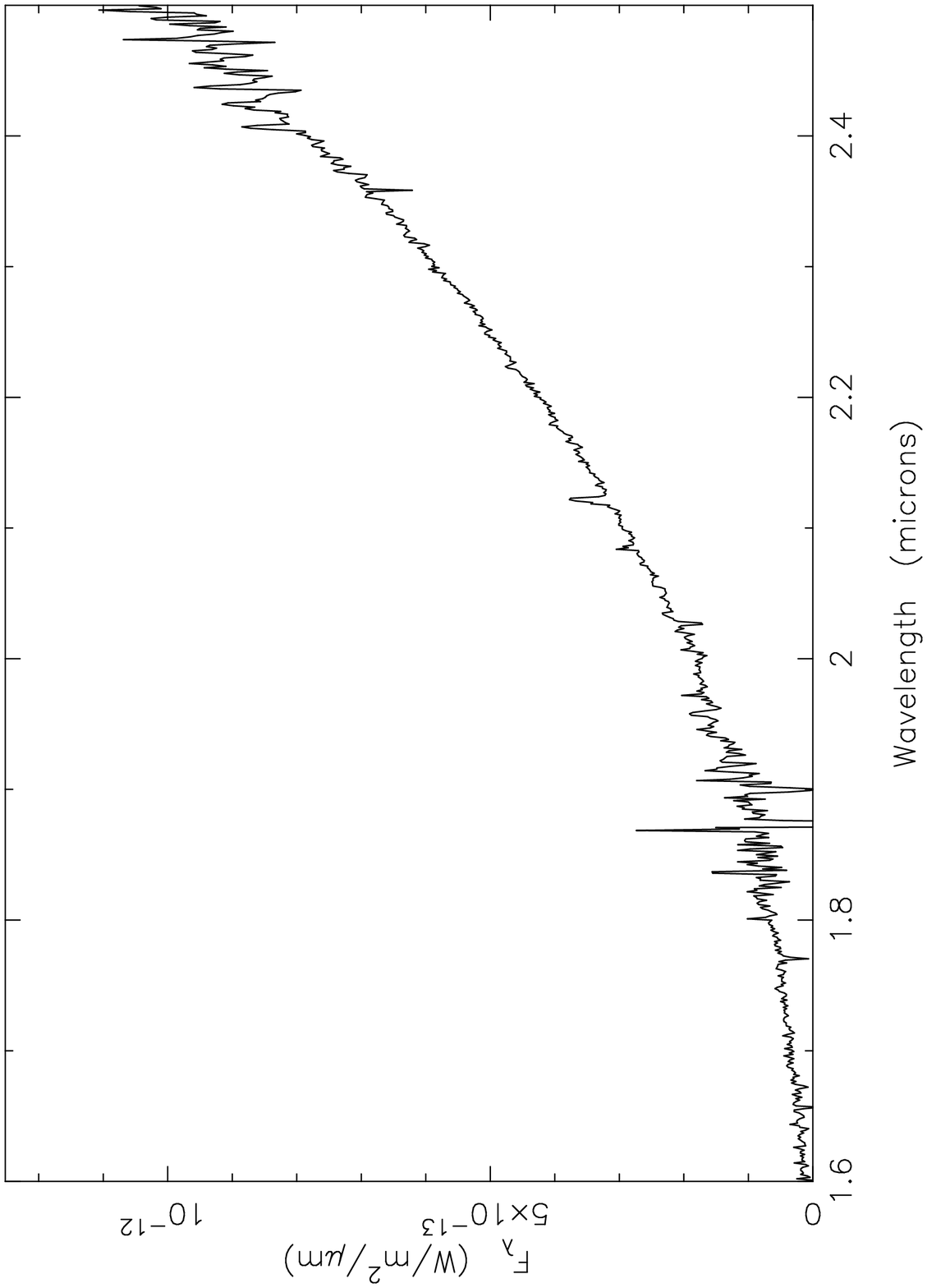}{3cm}{-90}{25}{25}{-20}{140}
\caption{a) Near-IR K-band image of one strong MYSO candidate showing a probably bipolar
nebula, extinction and associated cluster. b) Near-IR spectrum of the central source
in a). Note the very red, feature-less continuum characteristic of MYSOs. The 2.12$\mu$m
H$_{2}$ line is visible and this and other shock excited lines are strong along the
length of the outflow. The range 1.80$\mu$m$<\lambda<1.95\mu$m and $\lambda>2.39\mu$m are
affected by noise due to the atmosphere.}
\end{figure}

we are carrying out ground-based mid-IR
imaging
to check for extended and/or multiple point sources within the
18$^{\prime\prime}$ MSX beam.  
So far around 300 targets have been observed with about 10\%
revealed as extended sources, mostly likely compact H II regions or
PN. This is also the best way to identify MYSOs close to or superimposed
upon compact H II regions themselves. 

Near-IR K-band images of around 400 targets have been obtained to help 
identification and as a prelude to IR spectroscopy. About 80\%
are consistent with young star forming regions in that they show
nebulosity, clusters and/or extinction. Figure 2a shows the near-IR
image of one of our strong MYSO candidates. The near-IR spectrum
of this source (Figure 2b) is red and feature-less consistent with that
of many MYSOs. Our spectroscopy programme will also reveal any remaining
cooler post-MS objects that mimic MYSO characteristics in other ways.

\section{Conclusions}

The RMS survey is clearly finding large numbers of new MYSOs.  We
currently estimate that we will deliver a well-selected sample of
order 500 MYSOs with L$>10^{4}$L$_{\odot}$. This is an order of
magnitude larger than that currently known and will be close to being
complete for the galaxy at the higher luminosities. This number is
consistent with a simple estimate of the expected total number of
MYSOs in the galaxy based on the accepted star formation rate, an
assumed IMF and phase lifetime (Lumsden et al. 2002).

\end{document}